\documentclass[pdflatex,sn-mathphys-num]{sn-jnl}% Math and Physical Sciences Numbered Reference Style

\usepackage{graphicx}%
\usepackage{multirow}%
\usepackage{amsmath,amssymb,amsfonts}%
\usepackage{amsthm}%
\usepackage{mathrsfs}%
\usepackage[title]{appendix}%
\usepackage{xcolor}%
\usepackage{textcomp}%
\usepackage{manyfoot}%
\usepackage{booktabs}%
\usepackage{algorithm}%
\usepackage{algorithmicx}%
\usepackage{algpseudocode}%
\usepackage{listings}%

\theoremstyle{thmstyleone}%
\theoremstyle{thmstyletwo}%

\theoremstyle{thmstylethree}%

\usepackage[utf8]{inputenc}
\usepackage[T1]{fontenc}
\usepackage{url}

\usepackage{nicefrac}
\usepackage{microtype}
\usepackage{array}

\begin{document}

\title[GABIF: Guided Antibody Inverse Folding]{GABIF: An Adaptive Guided Graph Denoising Framework for Antibody
Conditional Inverse Folding}

\author[1]{\fnm{Mingwei} \sur{Liu}}
\equalcont{These authors contributed equally to this work.}

\author[1]{\fnm{Xiao} \sur{Liu}}
\equalcont{These authors contributed equally to this work.}

\author[2]{\fnm{Kai} \sur{Yi}}
\author[3]{\fnm{Feng} \sur{Tan}}
\author[1]{\fnm{Yu Guang} \sur{Wang}}

\affil[1]{\orgname{Shanghai Jiao Tong University}}
\affil[2]{\orgname{MRC Laboratory of Molecular Biology, Cambridge CB2 0QH, UK}}
\affil[3]{\orgname{Merck}}

\abstract{Antibodies play a central role in the human immune system, and their design is critical for vaccine development, therapeutic intervention, and improving human longevity. However, the vast combinatorial search space of antibodies poses a fundamental challenge for \textit{de novo} discovery. Recent advances in deep learning, particularly denoising diffusion models, have demonstrated strong potential in protein inverse folding. Nevertheless, existing approaches are largely constrained to learning from natural antibody repertoires, thereby generating sequences that closely resemble natural antibodies. Such similarity, however, does not necessarily translate into improved functional performance in downstream applications.
In this work, we develop a novel generative protein design model that enables the generation and optimization of antibody sequences tailored to specific downstream tasks and target metrics. 
Besides training on given an antibody dataset, we guide the generative model by an antibody-antigen binding affinity predictor for conditional discrete diffusion. 
The model is pre-trained on a given antibody dataset. Meanwhile, an antibody-antigen binding affinity predictor is incorporated as a conditional signal in a discrete diffusion framework, where a regressor serves as the conditional probability to guide the learning of the amino acid probability distribution of sequences under the objective of affinity optimization. 
\textit{In silico} experiments demonstrate that our antibody design model achieves performance comparable to state-of-the-art approaches, while the proposed regressor effectively improves binding affinity, as validated by an independent external predictor. Furthermore, our results highlight that the combination of finetuning and supervised guidance provides a powerful paradigm for adapting pretrained models to downstream tasks.}


\maketitle

\section{Introduction}
Protein inverse folding aims to generate amino acid (AA) sequences that are compatible with a target tertiary structure. As protein function is fundamentally determined by structure, inverse folding has become a core component of modern computational protein design pipelines. In recent years, this paradigm has attracted significant attention in therapeutic antibody engineering, where the ability to generate functional antibody sequences from predefined structural scaffolds has the potential to substantially accelerate drug discovery and biomedical development. Rationally designed antibodies play a central role in a broad range of applications, including targeted cancer therapy, immune checkpoint modulation, antiviral therapeutics, and precision diagnostics. However, compared with general protein design, antibody design imposes substantially stricter requirements on sequence generation due to the highly specific biochemical, structural, and functional constraints associated with antigen recognition and binding.

Despite rapid progress in computational biology and deep learning, antibody design remains fundamentally challenging because of the enormous combinatorial search space of protein sequences. Traditional antibody engineering pipelines often rely heavily on expert knowledge and involve complex procedures such as antibody--antigen co-modeling, molecular dynamics simulations, and complementary-determining region (CDR) optimization. These approaches are computationally expensive, difficult to scale, and typically require substantial domain expertise. Recent advances in deep learning have therefore motivated the development of data-driven approaches for protein inverse folding. In particular, transformer-based protein language models (PLMs) and geometric deep learning frameworks have demonstrated strong capabilities in capturing sequence--structure relationships and generating biologically plausible protein sequences~\cite{jumper2021alphafold, alphafold3, alphabind, baek2021accurate}. Existing approaches mainly formulate inverse folding as a conditional generation problem, where the target protein structure serves as conditioning information for recovering compatible AA sequences.

Current computational protein design methods are largely dominated by masked language models (MLMs)~\cite{doi:10.1126/science.ade257, Madani2023}, autoregressive sequence models~\cite{Hsu2022.04.10.487779, NEURIPS2021_f51338d7, nijkamp2023progen2}, and more recently, diffusion probabilistic models. While MLMs and autoregressive models have achieved promising performance, their generation mechanisms are often constrained by local token prediction objectives or sequential decoding processes, which may limit the diversity and controllability of generated sequences. In contrast, diffusion models provide a flexible probabilistic framework capable of generating diverse and high-quality samples through iterative denoising. Their strong representation capacity and stable training dynamics have led to substantial success across various generative modeling tasks, including image synthesis, molecular generation, and protein design.

Several recent studies~\cite{diffab, rfantibody, iggm} have explored diffusion-based frameworks for antibody generation and inverse folding. Although these methods are capable of producing structurally plausible antibodies, they predominantly learn the distribution of naturally occurring antibody repertoires. Consequently, generated sequences tend to remain close to naturally observed antibodies rather than being explicitly optimized for downstream therapeutic objectives. However, high similarity to natural antibodies does not necessarily translate into improved functional performance in real-world applications. In practical therapeutic antibody development, downstream properties such as antigen-binding affinity, specificity, stability, and developability are often substantially more important than merely reproducing natural sequence distributions. Therefore, a key challenge in modern antibody generation lies in moving beyond natural-distribution imitation toward property-guided therapeutic optimization.

To address this challenge, we propose \textbf{GABIF}, a guided-regressor generative framework for controllable antibody design and property-aware protein inverse folding. Rather than merely modeling the distribution of naturally occurring antibodies, GABIF enables task-oriented sequence generation conditioned on downstream objectives such as binding affinity optimization. The proposed framework adopts a discrete diffusion-based inverse folding architecture as the generative backbone and can be pretrained on arbitrary protein or antibody datasets to learn sequence--structure relationships in a data-driven manner. Furthermore, GABIF is designed as a modular guided generation framework that can flexibly incorporate external property predictors during the sampling process. In this work, we integrate the antibody--antigen binding affinity predictor ProAffinity-GNN~\cite{zhou2024proaffinity} as a guidance regressor to steer the diffusion trajectory toward antibody sequences with enhanced binding affinity. This design allows GABIF to unify pretrained generative modeling and downstream property optimization within a single controllable generation framework.
Inspired by classifier-guided diffusion and Bayesian posterior sampling, the affinity predictor provides conditional signals that steer the generation trajectory toward antibody sequences with desirable binding properties. This guided generation mechanism enables GABIF to move beyond passive modeling of natural antibody distributions and instead perform task-oriented optimization during sequence generation.

Extensive \textit{in silico} experiments demonstrate that GABIF achieves competitive or superior performance compared with state-of-the-art antibody inverse folding approaches across multiple evaluation benchmarks, including sequence recovery, structural consistency, and binding affinity estimation. Furthermore, ablation studies verify the effectiveness of the proposed guidance strategy and highlight the importance of supervised conditional generation for downstream therapeutic optimization. Our results suggest that the synergy between pretrained generative models and property-guided diffusion provides a powerful paradigm for controllable antibody design and broader AI-driven therapeutic discovery.

Our main contributions are summarized as follows:
\begin{itemize}
\item We propose GABIF, a guided discrete diffusion framework for antibody inverse folding with conditional affinity optimization.
\item We integrate an external affinity predictor into the diffusion sampling process to steer generation toward therapeutically desirable antibody sequences.
\item Extensive experiments demonstrate that GABIF achieves strong structural fidelity while substantially improving predicted binding affinity compared with existing approaches.
\end{itemize}

Code and pretrained models are publicly available at \url{https://github.com/Mintanxon/GABIF}.

\section{Related Work}

Denoising diffusion models have recently emerged as a powerful class of generative frameworks due to their ability to produce high-quality and diverse samples across multiple modalities~\citep{sohl-dickstein2015deep, DDPM, chen2021wavegrad}. Originating from stochastic diffusion processes and score-based generative modeling, diffusion probabilistic models iteratively transform noise into structured data and have demonstrated strong performance in image, waveform, and molecular generation tasks. Beyond continuous domains, substantial efforts have extended diffusion models to discrete spaces for modeling categorical and sequential data~\citep{mittal2021symbolicdiffusion, seff2019discrete, vignac2023digress}. In particular, discrete diffusion frameworks have shown notable advantages in preserving the structural properties of graph and sequence data, making them highly suitable for biological applications.

In computational protein and antibody design, recent advances in protein language models (PLMs) and geometric deep learning have significantly improved the ability to generate biologically plausible protein sequences~\citep{alley2019unified, shin2021protein, saka2021antibody, akbar2022fitforpurpose}. More recently, diffusion-based approaches such as DiffAb, AntiFold, and IgGM~\citep{antifold, iggm, diffab} have explored antibody generation and inverse folding using denoising diffusion mechanisms. While these methods achieve promising structural fidelity and controllability, they primarily focus on modeling natural antibody distributions and lack explicit optimization toward downstream therapeutic objectives such as binding affinity. In contrast, our work introduces a guided-regressor diffusion framework that integrates external property predictors into the generation process, enabling controllable and affinity-aware antibody design.

\section{Methods}
\begin{figure}[htbp]
\centering
\includegraphics[width=0.95\textwidth]{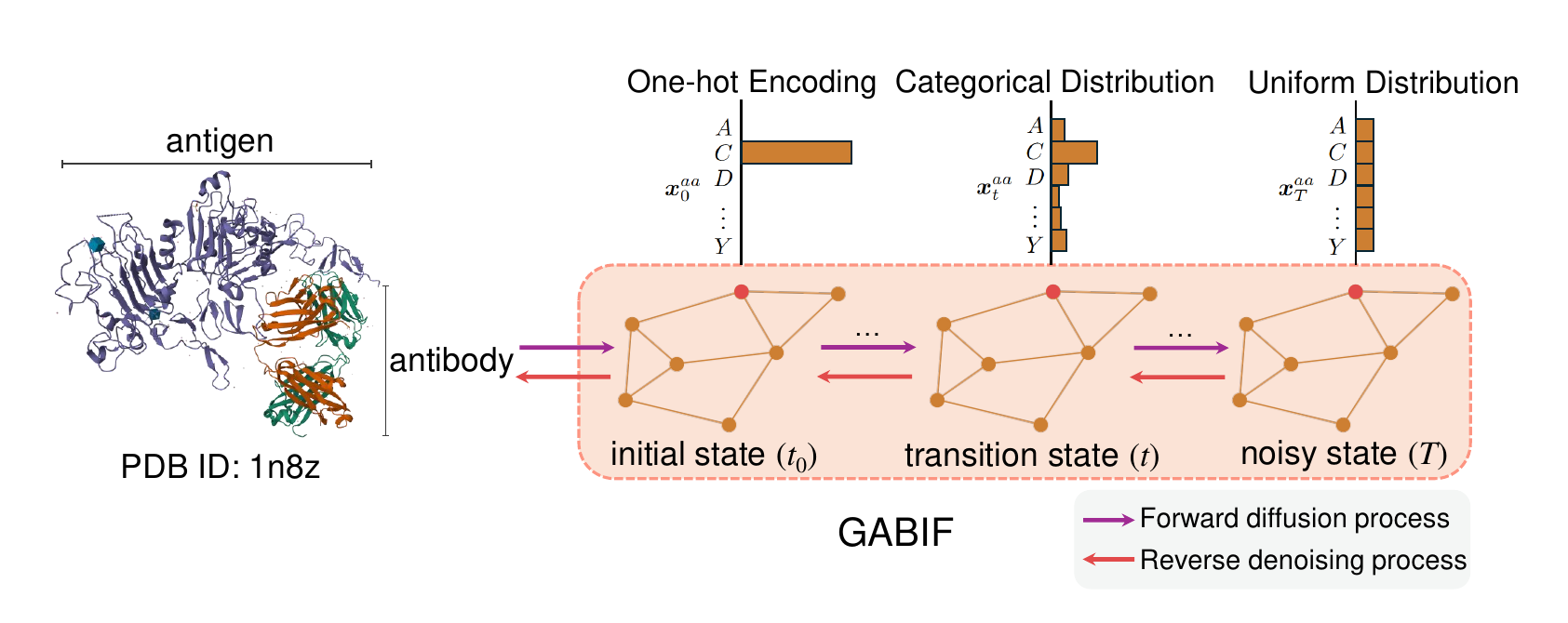}
\caption{Overview of the diffusion and denoising process in GABIF. Given an antigen--antibody complex structure, the model first constructs a residue-level graph representation. During the forward diffusion process, noise is progressively added to the amino acid (AA) types, transforming one-hot encoded sequences into a fully noisy uniform distribution. In the reverse process, sequences are iteratively denoised from sampled noise using the diffusion denoising network. Finally, an \textit{argmax} operation converts the predicted probability distributions into one-hot vectors, yielding the generated antibody amino acid sequences.}
\label{fig1}
\end{figure}

GABIF is derived from GraDe-IF \cite{gradeif} and operates as a graph neural network-based diffusion model for protein inverse folding.
We represent the residue graph of a protein as $\mathcal{G}=(\boldsymbol{X}, \boldsymbol{E})$.
Each AA in the protein corresponds to a node in the residue graph.
$\boldsymbol{X} \in \mathbb{R} ^{n\times m}$ denotes the node matrix, where $n$ is the number of residues, and $m$ represents the dimensionality of residue attributes, including AA type $\boldsymbol{X}^{\rm aa}\in \{0,1\}^{|\mathcal{A}|}$ (one-hot encoded, where $\mathcal{A}$ refers to the alphabet), 3-dimensional coordinates position $\boldsymbol{X}^{\rm pos}$ and other properties like physicochemical attributes.
$\boldsymbol{E}$ denotes the edges of the residue graph, encoding contact relationships between AA nodes. Each edge is assigned with additional attributes, such as interatomic distances and peptide bond positions.
Our model performs diffusion-based generation on this residue graph to solve the protein inverse folding task.
Figure \ref{fig1} illustrates the overview of the diffusion and denoising process of GABIF based on antibody structures.
\subsection{Protein Inverse Folding Problem}
In this work, protein inverse folding is defined as a conditional generation problem: Given a residue graph $\mathcal{G}$ encoding a protein's backbone structure (represented by $\boldsymbol{X}^{pos}$), the objective is to generate novel AA sequences which could fold into the given structure.
More specifically, the inverse folding problem is formalized as learning the conditional distribution $p(\boldsymbol{X}^{aa}|\boldsymbol{X}^{pos})$ from the native protein structures and AA sequences.
During training, the model is required to recover the native AA sequence from the given protein structure by minimizing the cross-entropy loss between predicted and native AA probabilities.
During inference, the model samples novel AA sequences from the learned distributions, thereby enabling \textit{de novo} protein design.

\subsection{Discrete Diffusion Process in Inverse Folding}
Denoising Diffusion Probabilistic Models (DDPM) have established remarkable success in image generation, by implementing a Markov chain that gradually adds Gaussian noise to corrupt training data, then reverses this process and samples from Gaussian distribution for generation \cite{DDPM}.

Due to the categorical nature of one-hot encoded AA sequences, we adopt the diffusion models for discrete data, Discrete Denoising Diffusion Probabilistic Models (D3PM) \cite{D3PM} to define diffusion process on the discrete state space.
The basic idea of D3PM is to corrupt one-hot encoded AA labels into the noise distribution not Gaussian distribution, like uniform distribution.
Specifically, for the diffusion process in inverse folding, we gradually add noise to each AA node $x^{aa}$ by transition matrix $\boldsymbol{Q}_t$.
We formally define the diffusion process as:
\begin{equation}
q(\boldsymbol{x}^{aa}_t|\boldsymbol{x}^{aa}_{t-1})=\text{Categorical}(\boldsymbol{x}_t^{aa}; \boldsymbol{x}_{t-1}^{aa}\boldsymbol Q_t)
\end{equation}
where:
\begin{itemize}
    \item $t$ denotes the step of diffusion process.
    \item $[\boldsymbol{Q}_t]_{ij}=q(x_t=j|x_{t-1}=i)$ is the time-dependent transition matrix:
    \begin{equation}\label{qt_uniform}
        \boldsymbol{Q}_t=\alpha_t\boldsymbol{I}+(1-\alpha_t)\frac{1}{|\mathcal{A}|}\boldsymbol{1}\boldsymbol{1}^\top
    \end{equation}
    \item $\alpha_t$ is the annealing schedule preserving original identity and undergoes a progressive decay until 0 as $t \to \infty$.
\end{itemize}
The final stationary distribution could be defined by:
\begin{equation}
\lim_{t\to \infty}q(\boldsymbol{x}_t)=\text{Uniform}(\mathcal{A})
\end{equation}
For Equation \ref{qt_uniform}, one could replace the uniform noise into biased noise by incorporating biological priors:
\begin{equation}
    \boldsymbol{Q}_t=\alpha_t\boldsymbol{I}+(1-\alpha_t)\boldsymbol{M}
\end{equation}
where $\boldsymbol{M}$ denotes the biased noise (e.g., BLOSUM substitution probabilities calculated from natural protein database).
Consequently, the marginal probability distribution at any step $t$ could be derived through the cumulative product of transition matrices:
\begin{equation}\label{qt_cumulative}
q(\boldsymbol{x}^{aa}_t|\boldsymbol{x}^{aa}_{0})=\text{Categorical}(\boldsymbol{x}_t^{aa}; \boldsymbol{x}_{0}^{aa}\bar{\boldsymbol Q}_t)
\end{equation}
where $\bar{\boldsymbol Q}_t=\boldsymbol{Q}_1\dots \boldsymbol{Q}_t=\prod\limits_{s=1}^{t}\boldsymbol{Q}_s$.
\subsection{Reverse Denoising Process}
The diffusion process corrupts the original data into noise through a forward Markov chain, while the reverse process progressively denoises to generate sample data.
If the true reverse distributions $q(\boldsymbol{x}^{aa}_{t-1}|\boldsymbol{x}^{aa}_{t})$ are known, one could reconstruct data by iteratively denoising from random noise $\boldsymbol{x}^{aa}_{t} \sim \text{Uniform}(\mathcal{A})$.
The objective of diffusion models is to train denoising models to approximate the true reverse distributions through parameterized neural networks $p_\theta(\boldsymbol{x}^{aa}_{t-1}|\boldsymbol{x}^{aa}_{t})$.
From D3PM, the posterior distribution $q(\boldsymbol{x}^{aa}_{t-1}|\boldsymbol{x}^{aa}_{t})$ given $\boldsymbol{x}_0^{aa}$ is:
\begin{equation}
q(\boldsymbol{x}^{aa}_{t-1}|\boldsymbol{x}^{aa}_{t},\boldsymbol{x}^{aa}_{0})=\text{Cat}\left(\boldsymbol{x}^{aa}_{t-1};\frac{\boldsymbol{x}^{aa}_{t}\boldsymbol Q_t^\top\odot\boldsymbol{x}^{aa}_{0}\bar{\boldsymbol Q}_{t-1}}{\boldsymbol{x}^{aa}_{0}\bar{\boldsymbol Q}_{t}(\boldsymbol{x}^{aa}_{t})^\top}\right)
\end{equation}
where $\text{Cat}(x^{aa})$ is short for categorical distribution $\text{Categorical}(x^{aa})$.
The reverse process is parameterized to predict the logits of the distribution:
\begin{equation}\label{eq:denoise}
p_\theta(\boldsymbol{x}^{aa}_{t-1}|\boldsymbol{x}^{aa}_{t})\propto \sum_{\tilde {\boldsymbol{x}}^{aa}_{0}}q(\boldsymbol{x}^{aa}_{t-1}|\boldsymbol{x}^{aa}_{t},\tilde{\boldsymbol{x}}^{aa}_{0})\tilde{p}_\theta(\tilde{\boldsymbol{x}}^{aa}_{0}|\boldsymbol{x}^{aa}_{t})
\end{equation}
Since both distribution $q(\boldsymbol{x}^{aa}_{t-1}|\boldsymbol{x}^{aa}_{t})$ and $p_\theta(\boldsymbol{x}^{aa}_{t-1}|\boldsymbol{x}^{aa}_{t})$ factorize, the KL divergence between $q$ and $p_\theta$ could be easily summed for each random variable.
After training the denoising network, novel AA sequences could be designed by sampling from the noise with given native structure and denoising under multiple iterations.
\subsection{Conditional Inverse Folding}
Notably, inverse folding models alone could only generate novel \textbf{natural-protein-like} sequences by learning native AA probability distributions, but higher similarity to natural proteins does not guarantee better performance of generated proteins (e.g., higher binding affinity or improved thermostability) \cite{evolvepro}.
In typical protein design practice, a large number of candidate sequences are generated using inverse folding models, and subsequently screened to identify ideal mutants, which is often resource-intensive and time-consuming.
To accelerate this process, it is required to guide the generation process towards desired properties.

In this paper, we propose the concept of conditional inverse folding, where a regressor is incorporated as guidance during the diffusion process.
The concept of conditional inverse folding is motivated by classifier-guided conditional generation in diffusion models \cite{diffusion-guidance}.
In conditional generation, the model generates content $x$ given a condition $y$, i.e., it learns the conditional probability distribution $p(x|y)$.
By Bayes rules:
\begin{equation}\label{bayes}
p(x|y)=\frac{p(y|x)\cdot p(x)}{p(y)}
\end{equation}
The gradient of the logarithm of Equation \ref{bayes}:
\begin{equation}\label{guidance gradient}
\begin{aligned}
\nabla_x \log p(x|y)&=\nabla_x\log\frac{p(y|x)\cdot p(x)}{p(y)}\\
&=\nabla_x (\log p(x) + \log p(y|x) - \log p(y))\\
&=\nabla_x\log p(x) + \nabla_x \log p(y|x)
\end{aligned}
\end{equation}
The first term represents the gradient of the unconditional diffusion process, while the second term corresponds to the classifier gradient. A scaling parameter $\alpha$ is incorporated into Equation \ref{guidance gradient} to adjust the signal of classifier guidance:
\begin{equation}\label{eq:guidance gradient}
\nabla_x \log p_\alpha(x|y)=\nabla_x\log p(x) + \alpha\nabla_x \log p(y|x)
\end{equation}
This classifier-guided approach allows the model to perform conditional generation towards the label specified by the classifier.

In conditional inverse folding, an appropriate predictor is required for guidance according to the target properties. For antibody design, we employ an antibody-antigen binding affinity predictor to guide the model in generating novel antibodies with improved affinity.
\subsection{Final Step Guidance}
A challenge in classifier guidance is that, in the earlier steps of the diffusion process, the generated samples are highly noisy.
This noisy input could lead to inaccurate predictions by the classifier and potentially cause a divergence from the true labels.
One potential solution is to train a classifier that predicts the final labels from the noisy data. However, for antibody design, predicting the binding affinity without noise is still difficult, and predicting from noisy data is even more challenging.
Consequently, we adopt a final step guidance strategy, where guidance is applied not from the beginning, but after a series of denoising steps. This approach maintains the accuracy and stability of the predictions on low-noise data.
\subsection{Architecture of GABIF Denoising Networks}
\begin{figure}[htbp]
\centering
\includegraphics[width=0.8\textwidth]{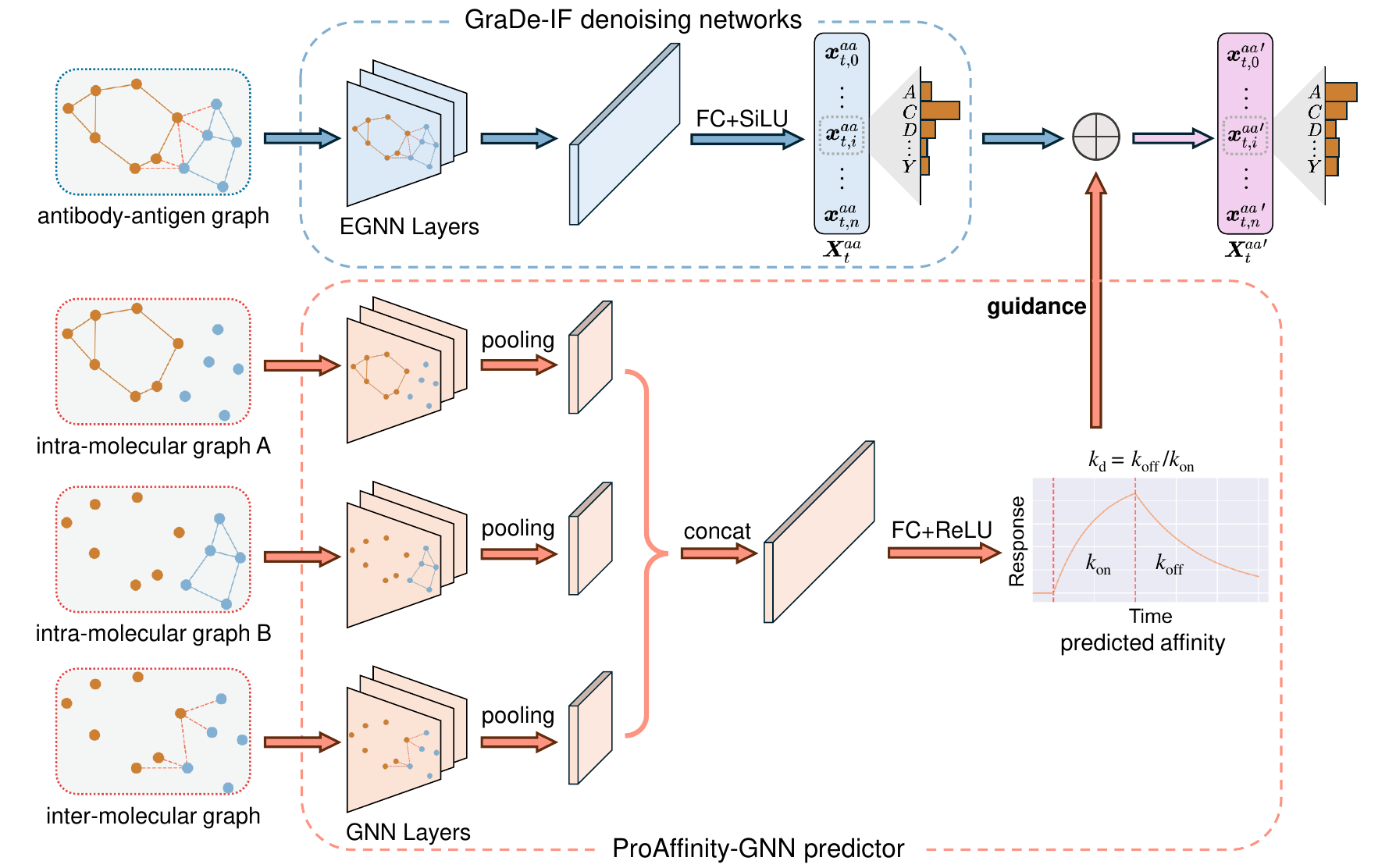}
\caption{The architecture of GABIF denoising networks guided with affinity regressor.}
\label{fig2}
\end{figure}

The full architecture of the GABIF denoising networks is depicted in Figure \ref{fig2}. The backbone of GABIF model is derived from GraDe-IF, which was originally trained on the general protein structure dataset CATH \cite{CATH}. To adapt it for antibody design tasks, we further finetune the model on antibody structures from SAbDab \cite{sabdab} to enable antibody design capabilities.
The model leverages the rotation equivariance of Equivariant Graph Denoising Network (EGNN) \cite{EGNN} to capture structural features of biomolecules in 3-dimensional space and integrates DDIM to accelerate the sampling process of the diffusion model. For the guidance in conditional inverse folding, we use ProAffinity-GNN as the binding affinity predictor. This module separately constructs intra-molecular and inter-molecular graphs, and utilizes GNN modules to predict the binding strength between two interacting molecules. ProAffinity-GNN is also originally designed for general protein complexes. Therefore, when integrating it into GABIF, we train the model using antibody-antigen complexes which include experimentally measured binding affinities from SAbDab. We show the detailed sampling and inference process in Appendix \ref{appendixa}. The model and datasets used in this paper are available at \url{https://github.com/Mintanxon/GABIF}.

\section{Experiments}
Our evaluation benchmark is divided into three parts: sequence evaluation metrics, structure evaluation metrics and binding affinity estimation \cite{antibody-design}. We conduct the experiments with an NVIDIA RTX 4090 GPU.
For sequence and structure evaluations (see metric definitions in Appendix \ref{appendixb}), we test our model by redesigning each CDR region of both the heavy and light chains of antibodies using the SAb-23H2-Ab dataset released in IgGM paper \cite{iggm}, and compare the performance with current structure-based antibody inverse folding models \cite{antifold, diffab, iggm}. For each antibody we design 10 mutants, and the experiments for each model take approximately several hours to complete.
It is important to note that, due to the lack of a standard dataset split, current antibody design models are often trained on the latest public dataset or proprietary antibody data with self-defined training and test split, which potentially causes data leakage. We have confirmed that the test data in our evaluation do not overlap with the training data of GABIF.
For binding affinity estimation, we utilize a third-party predictor trained on antibody-antigen affinity data for a specific antigen, and then use it to evaluate the affinity of 50 designed antibodies.
We also carry out an ablation study to discuss the effects of finetuning GraDe-IF on the antibody dataset and the impact of full step guidance against final step guidance.

\subsection{Sequence and Structure Evaluation Metrics}
The experimental results of designing all six CDRs on the benchmark dataset is presented in Table \ref{table:evalution}:
\begin{table}[htbp]
    \belowrulesep=0pt
    \aboverulesep=0pt
    \centering
    \caption{Evaluation metrics of antibody design models (numbers in \textbf{bold} denote the best performance)}
    \label{table:evalution}
    \setlength{\tabcolsep}{3pt}
    \begin{tabular*}{\linewidth}{@{\extracolsep{\fill}}c|l|rrrrrr|r@{}}
         \toprule
            \multirow{2}{*}{Metric} & \multirow{2}{*}{Model} & \multicolumn{6}{c|}{CDR} & \multirow{2}{*}{Total}\\
            \cmidrule{3-8}
                & & H1 & H2 & H3 & L1 & L2 & L3  \\
            \midrule
            \multirow{4}{*}{CAAR}
                & GABIF           & 0.750 & 0.718 & 0.541 & 0.698 & 0.708 & 0.592 & 0.668 \\
                & AntiFold             & 0.749 & 0.738 & 0.584 & 0.751 & 0.767 & 0.751 & 0.723 \\
                & DiffAb               & 0.773 & 0.794 & 0.565 & 0.698 & 0.527 & 0.702 & 0.677 \\
                & IgGM                 & 0.790 & 0.727 & 0.467 & 0.568 & 0.445 & 0.621 & 0.603 \\
            \midrule
            \multirow{4}{*}{\shortstack{$\text{MSR}\uparrow$}}
                & GABIF           & 0.841 & \textbf{0.909} & \textbf{1.000} & 0.841 & 0.705 & \textbf{1.000} & \textbf{0.883} \\
                & AntiFold             & \textbf{0.886} & \textbf{0.909} & \textbf{1.000} & 0.818 & 0.523 & 0.886 & 0.837 \\
                & DiffAb               & 0.818 & 0.750 & 0.955 & 0.818 & 0.727 & 0.909 & 0.830 \\
                & IgGM                 & 0.795 & 0.864 & 0.955 & \textbf{0.932} & \textbf{0.795} & 0.955 & \textbf{0.883} \\
            \midrule
            \multirow{4}{*}{\shortstack{$\text{RMSD}\downarrow$ \\ WT:10.043}}
                & GABIF           & 12.211 & \textbf{12.248} & \textbf{13.601} & 11.639 & 11.710 & 11.946 & \textbf{12.226} \\
                & AntiFold             & 13.216 & 13.824 & 14.150 & 12.820 & 11.610 & \textbf{11.270} & 12.815 \\
                & DiffAb               & \textbf{11.995} & 13.760 & 14.318 & 12.288 & 11.998 & 13.383 & 12.957 \\
                & IgGM                 & 12.549 & 13.214 & 15.571 & \textbf{11.108} & \textbf{11.256} & 12.255 & 12.659 \\
            \midrule
            \multirow{4}{*}{\shortstack{$\text{TM-score}\uparrow$ \\ WT:0.800}}
                & GABIF          & 0.748 & \textbf{0.752} & \textbf{0.721} & \textbf{0.766} & 0.764 & 0.765 & \textbf{0.753} \\
                & AntiFold            & \textbf{0.762} & 0.728 & 0.712 & 0.755 & 0.765 & \textbf{0.781} & 0.750 \\
                & DiffAb              & 0.756 & 0.732 & 0.703 & 0.764 & \textbf{0.776} & 0.745 & 0.746 \\
                & IgGM                & 0.743 & 0.740 & 0.694 & 0.766 & 0.773 & 0.753 & 0.745 \\
            \midrule
            \multirow{4}{*}{\shortstack{$\text{lDDT}\uparrow$ \\ WT:0.855}}
                & GABIF             & 0.839 & 0.840 & 0.824 & 0.843 & 0.842 & 0.836 & 0.837 \\
                & AntiFold               & 0.844 & 0.837 & \textbf{0.827} & \textbf{0.845} & 0.846 & \textbf{0.842} & \textbf{0.840} \\
                & DiffAb                 & \textbf{0.845} & \textbf{0.840} & 0.824 & 0.842 & \textbf{0.847} & 0.839 & 0.840 \\
                & IgGM                   & 0.842 & 0.836 & 0.817 & 0.840 & 0.845 & 0.831 & 0.835 \\
            \midrule
            \multirow{4}{*}{\shortstack{$\text{DockQ}\uparrow$ \\ WT:0.657}}
                & GABIF             & 0.565 & \textbf{0.585} & \textbf{0.518} & \textbf{0.614} & 0.605 & 0.593 & 0.580 \\
                & AntiFold               & \textbf{0.597} & 0.569 & 0.517 & 0.605 & 0.612 & \textbf{0.616} & \textbf{0.586} \\
                & DiffAb                 & 0.595 & 0.565 & 0.512 & 0.612 & \textbf{0.624} & 0.582 & 0.582 \\
                & IgGM                   & 0.576 & 0.561 & 0.482 & 0.600 & 0.611 & 0.568 & 0.566 \\
            \bottomrule
        \end{tabular*}
\end{table}

For sequence evaluation, the evaluation metrics include CAAR and MSR. Notably, AAR reflects the similarity between the denoised output and the original sequence. During training, a higher AAR indicates that the model better captures the true AA probability distribution. However, during inference, an overly high AAR could reduce sampling diversity and may even fail to generate novel sequences. Therefore, an appropriate trade-off between fidelity and diversity for AAR is essential at the inference stage. We report both CAAR and MSR in our sequence evaluation metrics. As shown in Table \ref{table:evalution}, CAAR and MSR exhibit a negative correlation. While GABIF reports slightly lower CAAR compared to AntiFold and DiffAb, it outperforms IgGM and achieves the highest MSR. This indicates that GABIF achieves a good balance between fidelity and diversity.

The inverse folding models generate novel antibody sequences from the given structure. To assess the structural quality of the designed antibodies, we apply AlphaFold3 \cite{alphafold3} to predict the antigen-antibody complex structures. As a control, we also predict the complex structures for the native wild-type (WT) sequences, ensuring a consistent AlphaFold3-involved evalution setup. To ensure the reliability of AlphaFold3 predictions, WT antibodies with an lDDT score above 0.8 are retained for evaluation.

As shown in Table \ref{table:evalution}, GABIF achieves the lowest RMSD and highest TM-score. AntiFold exhibits the best performance on lDDT and DockQ, with GABIF is slightly behind. Besides, GABIF surpasses IgGM across all evaluation metrics, with both models verified to have no risk of data leakage.

When further comparing each individual CDR, GABIF achieves superior results on CDR-H2 and CDR-H3 of the heavy chain. CDR-H3 is widely considered as the critical region for determining antibody specificity and tends to be highly variable \cite{CDR}. Consequently, the superior performance of GABIF on CDR-H3 region highlights its practical applicability in antibody design.
\subsection{Binding Affinity Estimation}
\begin{figure}[htbp]
\centering
\includegraphics[width=0.6\columnwidth]{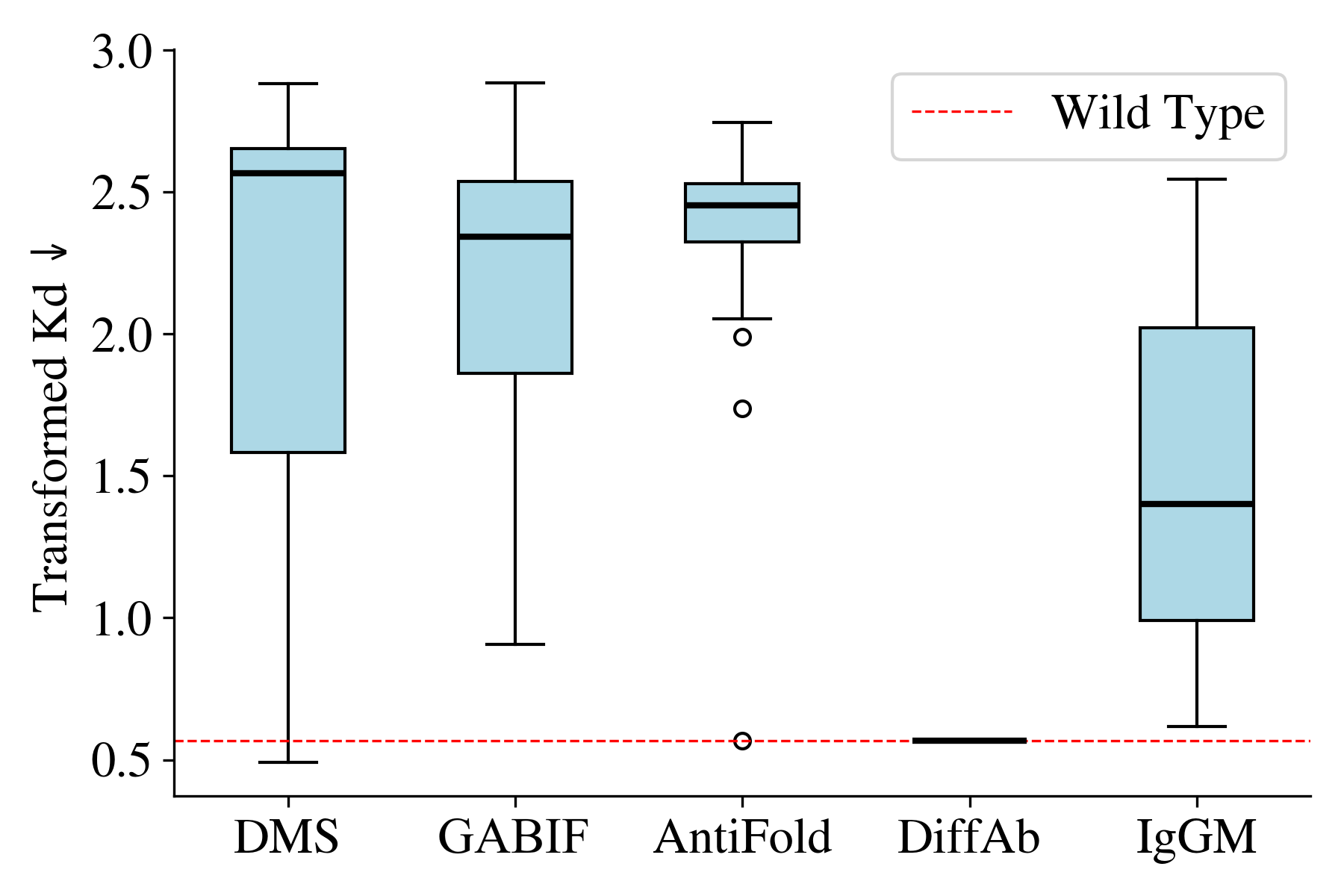}
\caption{The AlphaBind predicted affinities of HER2 antibody CDR-H3 mutants designed by different models. The red dashed line shows the binding affinity of the wild-type antibody.}
\label{fig3}
\end{figure}

Previous works on computational antibody evaluation primarily limited to sequence and structure evaluation metrics. In our work, we not only introduce regressor guidance for better affinity, but also contribute binding affinity evaluation methods. For $\textit{in silico}$ binding affinity estimation, we select an external predictor AlphaBind \cite{alphabind} to evaluate the binding affinity of the designed antibodies. Compared to training a general affinity predictor, it is easier and more accurate to train an antigen-specific affinity predictor. Consequently, we utilize the DMS dataset provided by AlphaBind, which includes more than 36,000 affinity values for HER2-specific antibody (PDB ID: 1n8z) CDR-H3 mutants \cite{alphabind-dataset}, to train the model as a judge. This model is then used to evaluate the affinities of 50 CDR-H3 mutants designed by each model.
The binding affinity is represented by transformed Kd in Alphabind. A lower Kd value indicates a stronger binding affinity. The predicted affinities are presented in Figure \ref{fig3}, where the red dashed line represents the wild-type antibody affinity. Data point overlapping with this line indicates that the designed sequence is identical to the native sequence, implying a failed design. As shown in Figure \ref{fig3}, DiffAb fails to design new sequences, and AntiFold also encounters design failures. Overall, the mutants designed by models demonstrate better affinity than the DMS mutants, suggesting that the potential of AI models in antibody design surpasses the traditional ration design. The mutants designed by IgGM exhibit the highest affinity, followed by GABIF and AntiFold. Considering the structural evaluation results, the relatively poor folding ability of IgGM may lead to less accurate affinity predictions. In conclusion, GABIF performs well both in structure and binding affinity evaluation metrics.
\begin{table}[htbp]
    \belowrulesep=0pt
    \aboverulesep=0pt
    \caption{Evaluation metrics for the ablation study. Numbers in \textbf{bold} indicate the best performance. ``Final step'' denotes GABIF with guidance applied only at the final denoising step, ``full step'' denotes GABIF with guidance throughout the entire diffusion process, ``finetune'' denotes the finetuned GABIF model without guidance, and ``original'' refers to the original GraDe-IF model.}
    \centering
    \label{table:ablation study}
    \setlength{\tabcolsep}{3pt}
    \begin{tabular*}{\linewidth}{@{\extracolsep{\fill}}c|l|rrrrrr|r@{}}
         \toprule
            \multirow{2}{*}{Metric} & \multirow{2}{*}{Model} & \multicolumn{6}{c|}{CDR} & \multirow{2}{*}{Total}\\
            \cmidrule{3-8}
                & & H1 & H2 & H3 & L1 & L2 & L3  \\
            \midrule
            \multirow{4}{*}{CAAR}
                & final step  & 0.750 & 0.718 & 0.541 & 0.698 & 0.708 & 0.592 & 0.668 \\
                & full step   & 0.706 & 0.657 & 0.519 & 0.656 & 0.694 & 0.573 & 0.634 \\
                & finetune            & 0.777 & 0.730 & 0.579 & 0.682 & 0.682 & 0.608 & 0.677 \\
                & original       & 0.550 & 0.542 & 0.413 & 0.401 & 0.523 & 0.355 & 0.464 \\
            \midrule
            \multirow{4}{*}{\shortstack{$\text{MSR}\uparrow$}}
                & final step  & 0.841 & 0.909 & \textbf{1.000} & 0.841 & 0.705 & \textbf{1.000} & 0.883 \\
                & full step   & 0.886 & 0.977 & \textbf{1.000} & 0.864 & 0.727 & 0.977 & 0.905 \\
                & finetune            & 0.818 & 0.841 & \textbf{1.000} & 0.841 & 0.705 & \textbf{1.000} & 0.867 \\
                & original       & \textbf{1.000} & \textbf{1.000} & \textbf{1.000} & \textbf{0.977} & \textbf{0.864} & \textbf{1.000} & \textbf{0.973} \\
            \midrule
            \multirow{4}{*}{$\Delta \text{pKa}\downarrow$}
                & final step   & -0.007 & -0.004 & -0.008 & -0.007 & -0.001 & -0.003 & -0.005 \\
                & full step    & \textbf{-0.009} & \textbf{-0.005} & \textbf{-0.028} & \textbf{-0.008} & \textbf{-0.003} & \textbf{-0.007} & \textbf{-0.010} \\
                & finetune             & -0.004 & -0.003 &  0.027 & -0.004 &  0.001 &  0.004 &  0.004 \\
                & original        &  0.023 &  0.013 &  0.024 &  0.013 &  0.000 &  0.025 &  0.016 \\
            \midrule
            \multirow{4}{*}{\shortstack{$\text{RMSD}\downarrow$ \\ WT:10.043}}
                & final step   & \textbf{12.211} & 12.248 & 13.601 & \textbf{11.639} & 11.710 & \textbf{11.946} & \textbf{12.226} \\
                & full step    & 12.896 & 12.234 & 13.688 & 13.287 & 11.672 & 12.634 & 12.735 \\
                & finetune             & 12.980 & 12.175 & 13.749 & 12.204 & 11.342 & 12.297 & 12.458 \\
                & original        & 13.119 & \textbf{12.071} & \textbf{12.676} & 12.233 & \textbf{10.711} & 14.301 & 12.519 \\
            \midrule
            \multirow{4}{*}{\shortstack{$\text{TM-score}\uparrow$ \\ WT:0.800}}
                & final step   & 0.748 & 0.752 & \textbf{0.721} & 0.766 & 0.764 & 0.765 & \textbf{0.753} \\
                & full step    & 0.746 & \textbf{0.754} & 0.714 & 0.746 & 0.767 & 0.755 & 0.747 \\
                & finetune             & 0.742 & 0.747 & 0.714 & \textbf{0.768} & 0.769 & 0.756 & 0.749 \\
                & original        & \textbf{0.750} & 0.746 & 0.698 & 0.759 & \textbf{0.781} & \textbf{0.767} & 0.745 \\
            \midrule
            \multirow{4}{*}{\shortstack{$\text{lDDT}\uparrow$ \\ WT:0.855}}
                & final step   & 0.839 & 0.840 & 0.824 & \textbf{0.843} & 0.842 & \textbf{0.836} & \textbf{0.837} \\
                & full step    & \textbf{0.841} & \textbf{0.842} & 0.822 & 0.839 & 0.843 & 0.834 & 0.837 \\
                & finetune             & 0.839 & 0.838 & \textbf{0.825} & 0.841 & 0.842 & 0.836 & 0.837 \\
                & original        & 0.834 & 0.836 & 0.815 & 0.839 & \textbf{0.847} & 0.820 & 0.832 \\
            \midrule
            \multirow{4}{*}{\shortstack{$\text{DockQ}\uparrow$ \\ WT:0.657}}
                & final step   & 0.565 & 0.585 & 0.518 & \textbf{0.614} & 0.605 & \textbf{0.593} & \textbf{0.580} \\
                & full step    & \textbf{0.572} & \textbf{0.594} & 0.502 & 0.589 & 0.609 & 0.583 & 0.575 \\
                & finetune             & 0.560 & 0.577 & \textbf{0.521} & 0.610 & 0.608 & 0.579 & 0.576 \\
                & original        & 0.565 & 0.562 & 0.499 & 0.597 & \textbf{0.628} & 0.526 & 0.563 \\
            \bottomrule
        \end{tabular*}
\end{table}
\subsection{Ablation Study}
To investigate the impact of various components of GABIF, we conduct ablation experiments as shown in Table \ref{table:ablation study} and Figure \ref{fig4} (see in Appendix \ref{appendixc}). From Table \ref{table:ablation study}, it is evident that the original GraDe-IF has a much lower CAAR than the others, suggesting that general protein inverse folding models perform poorly on antibody data. After finetuning on the antibody dataset, the CAAR improves significantly. The model with guidance shows a slight decrease in CAAR compared to the model without guidance, which indicates that the guidance could help to generate less natural-antibody-like sequences. The affinities predicted by ProAffinity-GNN are represented by pKa, where $\text{pKa} = -\log \text{Ka}=\log \text{Kd}$ and $\Delta \text{pKa}=\text{pKa}(\text{Mutant})-\text{pKa}(\text{WT})$. Therefore, the smaller the $\Delta \text{pKa}$, the greater the decrease in Kd, and the greater the improvement in affinity. When no guidance is applied, the mutant affinities are lower than the native antibody, while the affinity improves with the addition of ProAffinity-GNN guidance. However, the structure metrics of the full step guidance model are worse than those of the model without guidance, with the RMSD even exceeding the original GraDe-IF. In contrast, the final step version exhibits improved affinity and the best performance across all structure evaluations, confirming the conclusion above that GABIF excels in both structure and binding affinity metrics.

Figure \ref{fig4} presents the antibody affinities predicted by AlphaBind for different versions of GABIF. The affinities of all the model designs are higher than the DMS data. Similar to the $\Delta \text{pKa}$ in Table \ref{table:ablation study}, as the guidance interference increases, the Kd value gradually decreases. The consistent trend observed in both ProAffinity-GNN and AlphaBind validates the accuracy of \textit{in silico} affinity predictions.

In summary, both finetuning and final step guidance effectively enhance the model's performance, enabling GABIF to design antibody mutants with better performance and higher affinity.

\section{Conclusion}
In conclusion, we presented a guided-regressor diffusion framework for controllable antibody design and property-aware protein inverse folding. By integrating an external affinity predictor into the discrete diffusion sampling process, the proposed GABIF enables task-oriented antibody generation toward improved binding affinity while maintaining strong structural consistency. Experimental results demonstrate that GABIF achieves competitive performance against multiple state-of-the-art baselines. More broadly, our results suggest that combining pretrained generative modeling with guided diffusion provides a promising paradigm for controllable therapeutic protein design. In future work, we plan to extend the framework to broader biomolecular generation tasks, including enzyme engineering and multi-objective protein optimization.

\backmatter

\begin{appendices}

\section{Sampling Novel Antibodies}\label{appendixa}
The detailed algorithm of generating novel antibody sequences for GABIF is summarized as Algorithm \ref{alg:algorithm}. The model begins with sampling from the noise distribution and iteratively denoises the input by denoising networks. The binding affinity predictor ProAffinity-GNN predicts the affinity of generated antibody and calculates the gradient of higher affinity, thereby steering the antibody generation towards higher affinity. Notably, for final step guidance approach, the guidance is only applied at the final step.
\begin{algorithm}[htbp]
\caption{GABIF sampling with guidance}
\label{alg:algorithm}
\textbf{Parameter}: Guidance scaling parameter $\alpha$\\
\textbf{Output}: Generated antibody sequence $\boldsymbol{X}^{aa}_{0}$

\begin{algorithmic}[1]
\State Sample from noise distribution $\boldsymbol{X}^{aa}_{T}\sim \text{Uniform}(\mathcal{A})$.
\For{$t$ in $\{T, T-1, \dots, 1\}$}
\State Predict $\tilde p(\boldsymbol{X}^{aa}_{0}|\boldsymbol{X}^{aa}_{t})$ by denoising network.
\State Compute $p_\theta(\boldsymbol{X}^{aa}_{t-1}|\boldsymbol{X}^{aa}_{t})$ by Equation \ref{eq:denoise}.
\State Predict the affinity of $\boldsymbol{X}^{aa}_{t}$ by ProAffinity-GNN and compute the gradient of higher affinity.
\State Calculate guided  $\hat p_\theta(\boldsymbol{X}^{aa}_{t-1}|\boldsymbol{X}^{aa}_{t})$ by Equation \ref{eq:guidance gradient}.
\State Sample $\boldsymbol{X}^{aa}_{t-1}\sim \hat p_\theta(\boldsymbol{X}^{aa}_{t-1}|\boldsymbol{X}^{aa}_{t})$.
\EndFor
\State Sample $\boldsymbol{X}^{aa}_{0}\sim \hat p_\theta(\boldsymbol{X}^{aa}_{0}|\boldsymbol{X}^{aa}_{1})$.
\State \textbf{return} $\boldsymbol{X}^{aa}_{0}$
\end{algorithmic}
\end{algorithm}

\section{Definitions of Evaluation Metrics}\label{appendixb}
The metrics used in this paper are composed of sequence evaluation metrics and structure evaluation metrics.

Sequence evaluation metrics include:
\begin{itemize}
    \item CAAR: the recovery rate of AAs (AAR) in CDR regions.
    \item MSR: Mutation Success Rate, refers to the proportion of successful mutations which are not identical to the native sequence.
\end{itemize}
Structure evaluation is conducted using OpenStructure \cite{biasini2013openstructure}, which provides a convenient command-line interface (CLI) to measure various metrics between predicted and wild-type structures.
Structure evaluation metrics include:
\begin{itemize}
    \item RMSD: Root Mean Square Deviation, the average distance between corresponding atoms of two structures after optimal superposition. Lower RMSD values indicate higher structural similarity.
    \item TM-score \cite{tm-score}: Template Modeling Score, the global structural similarity between two protein structures normalized by protein length. It is less sensitive to local deviations and ranges from 0 to 1, where scores above 0.5 typically indicate correct folding \cite{tm-score-0.5}.
    \item lDDT \cite{mariani2013lddt}: Local Distance Difference Test, the local atomic-level accuracy by comparing the preservation of interatomic distances within a neighborhood. It is superposition-free and higher scores indicate better local structural accuracy.
    \item DockQ \cite{basu2016dockq}: the score to quantify the quality of biomolecule docking models by integrating metrics including interface RMSD (iRMSD), ligand RMSD (LRMSD), and fraction of retrieved native contacts (fnat). Higher scores reflect more accurate docking predictions.
\end{itemize}

\section{Binding Affinity Estimation in Ablation Study}\label{appendixc}
\begin{figure}[htbp]
\centering
\includegraphics[width=0.6\columnwidth]{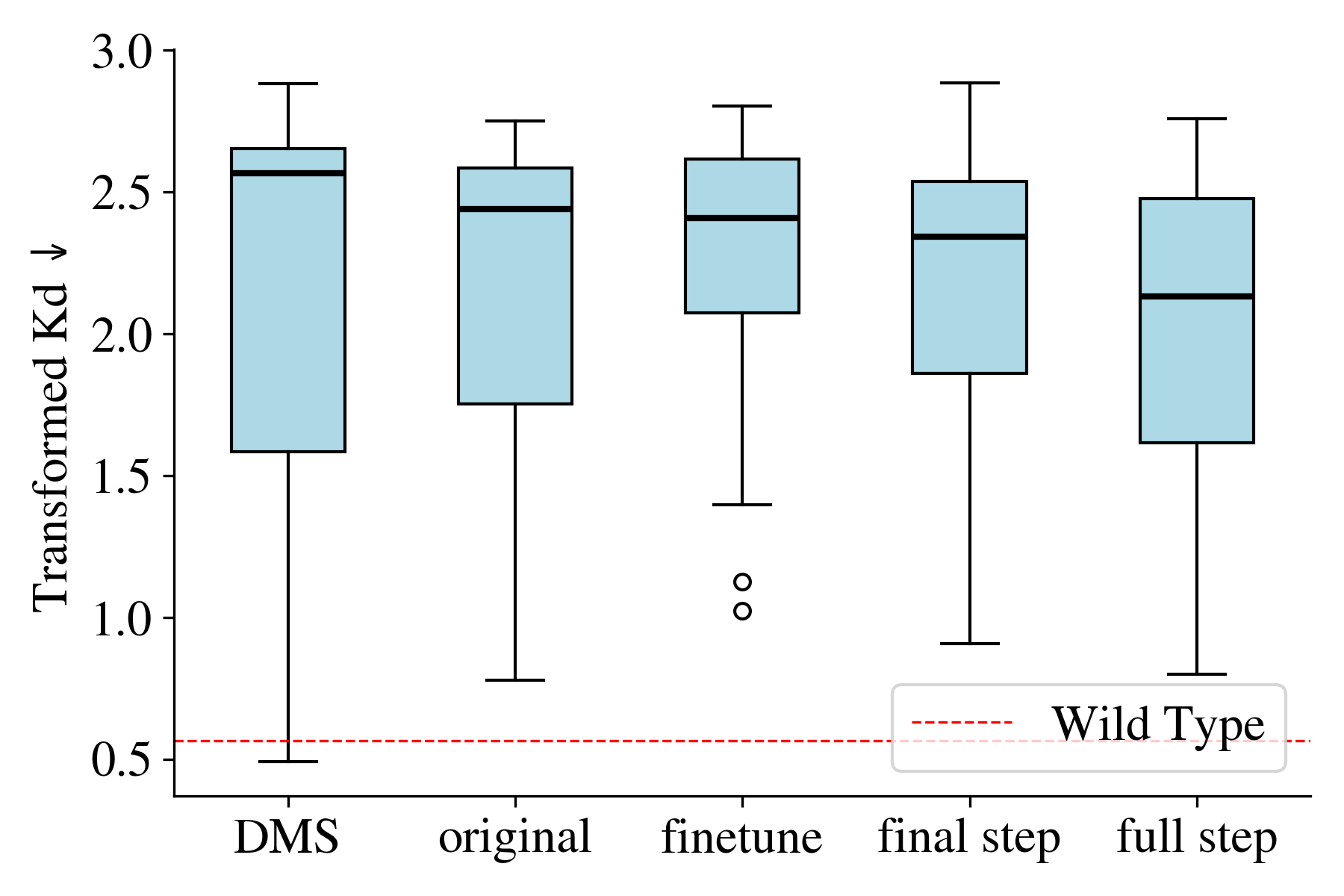}
\caption{The AlphaBind predicted affinities of HER2 antibody CDR-H3 mutants designed by GABIF in ablation study (``final step'' refers to GABIF model with final step guidance, ``full step'' refers to GABIF model with full step guidance, ``finetune'' refers to GABIF without guidance (finetuning only), ``original'' refers to the original GraDe-IF)}
\label{fig4}
\end{figure}

We test the binding affinities of different versions of GABIF sampled antibodies by AlphaBind to investigate the impacts of finetuning and guidance. The results are illstruated in Figure \ref{fig4}. The consistent trend of $\Delta \text{pKa}$ (Table \ref{table:ablation study}) and transformed Kd (Figure \ref{fig4}) predicted by different models validates the reliability of \textit{in silico} binding affinity estimation.

\end{appendices}

\bibliography{sn-bibliography}

\end{document}